\documentclass[twocolumn,showpacs,preprintnumbers,amsmath,amssymb]{revtex4}

\usepackage{bm,graphicx,dcolumn,bbm,epsfig}

\newcommand{\ds}{\displaystyle}
\newcommand{\cp}{{\mathbbm C}{\bf P}}
\newcommand{\tr}{\mbox{Tr}}

\newcommand{\be}{\begin{equation}}
\newcommand{\ee}{\end{equation}}
\newcommand{\ba}{\begin{array}}
\newcommand{\ea}{\end{array}}

\newcommand{\bra}[1]{\ensuremath{\langle #1 |}}
\newcommand{\ket}[1]{\ensuremath{| #1 \rangle}}
\newcommand{\abl}[2]{\ensuremath{\frac{\partial #1}{\partial #2}}}
\renewcommand{\ol}[2]{\ensuremath{\langle #1 | #2 \rangle}}
\newcommand{\matel}[3]{\ensuremath{\langle #1 | #2 | #3 \rangle}}

\begin{document}


\title{Wehrl entropy, Lieb conjecture and entanglement monotones}

\author{Florian Mintert$^{1,2}$ and Karol {\.Z}yczkowski$^{2,3}$}

\affiliation{
  $^1$Max Planck Institute for the physics of complex systems
  N\"othnitzerstr. 38
  01187 Dresden
}
\affiliation{
  $^2$Uniwersytet Jagiello\'nski,
  Instytut Fizyki im.~M.~Smoluchowskiego,
  ul.~Reymonta 4,  30-059 Krak\'ow, Poland}

\affiliation{
 $^3$Centrum Fizyki Teoretycznej, Polska Akademia Nauk,
  Al.~Lotnik{\'o}w 32/44, 02-668 Warszawa, Poland
}

\date{ October 23, 2003}

\begin{abstract}
We propose to quantify the entanglement of pure states
of $N \times N$ bipartite quantum systems by defining 
its Husimi distribution with respect to $SU(N)\times SU(N)$
coherent states. The Wehrl entropy is minimal if and only if 
the analyzed pure state is separable. The excess of the Wehrl entropy
is shown to be equal to the subentropy of the mixed state 
obtained by partial trace of the bipartite pure state.
This quantity, as well as the generalized (R{\'e}nyi) 
subentropies, are proved to be Schur-concave, 
so they are entanglement monotones and 
may be used as alternative measures of entanglement.
\end{abstract}

\pacs{03.67 Mn, 89.70.+c, }

\maketitle

\section{Introduction}

Investigating properties of a quantum state it is useful 
to analyze its phase space representation.
The {\sl Husimi distribution} is often very convenient to work with:
it exists for any quantum state, is non-negative for any
point  $\alpha$ of the classical phase space $\Omega$, and may be normalized 
as a probability distribution \cite{Hu40}.
 The Husimi distribution can be defined as the expectation 
value of the analyzed state $\varrho$
with respect to the coherent state $\ket{\alpha}$,
localized at the corresponding point $\alpha \in \Omega$.

If the classical phase space is equivalent to the plane one
uses the standard harmonic oscillator coherent states (CS),
but in general one may apply the group--theoretical
construction of Perelomov \cite{Pe86}.
For instance,  the group $SU(2)$ leads
to the spin coherent states parametrized by the sphere
$S^2=\cp^{1}$ \cite{Ra71,Gi72}, 
while the simplest, degenerated representation of 
$SU(K)$ leads to the higher vector coherent states
parametrized by points on the complex projective mani\-fold
$\cp^{K-1}$ \cite{He87,GS93,GK98}.
One may define the Husimi distribution of the 
$N$--dimensional pure state $|\phi\rangle$ 
with respect to $SU(K)$--CS for any $K \le N$,
but it is important to note that if $K=N$
any pure state is by definition $SU(N)$--coherent. 

Localization properties of a state under consideration
may be characterized by the {\sl Wehrl entropy},
defined as the continuous Boltzmann--Gibbs entropy 
of the Husimi function \cite{We78}.
The Wehrl entropy admits smallest values for coherent states,
which are as localized in the phase space, as allowed by the
Heisenberg uncertainty relation. Interestingly, this
property conjectured first by Wehrl \cite{We78} 
for the harmonic oscillator coherent states
was proved soon afterwords by Lieb \cite{Li78},
but the analogous result for the $SU(2)$--CS
still waits for a rigorous proof. This unproven property 
is known in the literature as the {\sl Lieb conjecture} \cite{Li78}.
It is known that the $SU(2)$--CS provide local minimum  of the Wehrl entropy 
\cite{Le88}, while a proof of the global minimum was given 
for pure states of dimension $N=3$ and $N=4$ only \cite{Sc79,Sc99}.
A generalized Lieb conjecture, concerning the R{\'e}nyi -Wehrl entropies
of integer order $q\ge 2$ occurred to be easier then the original
statement and its proof was given in \cite{GZ01,Sc99}.
It is also straightforward to formulate the Lieb conjecture
for Wehrl entropy computed with respect
to the $SU(K)$--coherent states \cite{SZ98,Su02},
but this conjecture seems not to be simpler 
than the original one.

In this work we consider the Husimi function
of an arbitrary  mixed quantum state $\rho$ of size $N$ with respect to
$SU(N)$ coherent states,
and calculate both its statistical moments and the Wehrl entropy.
The difference between the latter
quantity and the minimal entropy attained for pure states,
can be considered as a measure of the degree of mixing. 

Analyzing pure states of a bipartite system it is
helpful to define coherent states with respect to 
product groups. The Wehrl entropy of a given state
with respect to the $SU(N)\times SU(N)$--CS 
was first considered in the original paper of Wehrl \cite{We79}.
Recently Sugita proposed
to use moments of the Husimi distribution
defined with respect to such product coherent states
as a way to characterize the entanglement of 
the analyzed state \cite{Su03}.

In this paper we follow his idea
and compute explicitly the Wehrl entropy 
and the generalized R{\'e}nyi--Wehrl entropies
for any pure state $|\Psi\rangle$
of the $N \times N$ composite system.
The Husimi function is computed with respect to 
$SU(N)\times SU(N)$--CS, so the Wehrl entropies
achieve their minimum  if and only if
the state $|\Psi\rangle$ is a product state.
Hence the entropy excess defined as 
the difference with respect to the minimal value
quantifies, to what extend the analyzed state is entangled.
The Wehrl entropy excess is shown to be equal to the subentropy 
defined by Jozsa et al. \cite{JRW94}. Calculating the excess
of the R{\'e}nyi--Wehrl entropies 
we define the {\sl R{\'e}nyi subentropy} --
a natural continuous generalization of the subentropy.

Several different measures of quantum entanglement 
introduced in the literature 
(see e.g. \cite{VP98,Vi00,Ho02} and references therein)
satisfy the list of axioms formulated in \cite{VP98}.
In particular, quantum entanglement cannot increase 
under the action of any local operations, 
and a measure fulfilling this property is called 
entanglement monotone. 

The non-local properties of a pure state of 
an $N \times N$ system has to be characterized by 
$N-1$ independent monotones \cite{Vi00}.
In this work we demonstrate that $N-1$ 
statistical moments of properly defined Husimi functions naturally
provide such a set of parameters, since they 
are Schur--concave functions of the Schmidt coefficients and thus are non-increasing
under local operations.

\section{Husimi function and Wehrl entropy}

Any density matrix
$\varrho\in{\mathbbm C}^{N\times N}$
can be represented by its Husimi function $H_{\varrho}(\alpha)$,
that is defined as its expectation value with respect to
coherent states $\ket{\alpha}$,
\be
H_{\varrho}(\alpha)=\matel{\alpha}{\varrho}{\alpha}
\ee

In general one may define the Husimi distribution
with respect to $SU(K)$-coherent states with  $2\le K\le N$,
but most often one computes the Husimi distribution
with respect to the $SU(2)$ coherent states, also
called spin coherent states \cite{Ra71,Gi72}.
Let us recall that any family of coherent states $\{|\alpha\rangle \}$
has to satisfy the resolution of identity
\begin{equation}
\int_{\Omega} d\mu(\alpha)\ket{\alpha}\bra{\alpha}={\mathbbm 1} \ ,
\label{resolution}
\end{equation}
where $d\mu(\alpha)$ is a uniform measure on the classical mani\-fold
$\Omega$.
In the case of the degenerated representation of $SU(K)$ 
coherent states \cite{He87,GS93,GK98}
this manifold,
$\Omega_K=U(K)/[U(K-1)\times U(1)]={\mathbbm C}{\bf P}^{K-1}$,
is just equivalent to the space of all 
pure states of size $K$.
This complex projective space 
arises from the $K$--dimensional Hilbert
space ${\cal H}_K$ by taking into account normalized states,
and identifying all elements of ${\cal H}_K$,
which differ by an overall phase only. 
 In the simplest case
of $SU(2)$ coherent states it is just the well known 
Bloch sphere,
$\Omega_2={\mathbbm C}{\bf P}^{1}=S^2$.

Analyzing a mixed state
of dimensionality $N$ we are going to use
$SU(N)$ coherent states,  which will be denoted by
$|\alpha_N\rangle$.
With this convention the space $\Omega$
is equivalent to the complex projective space $\cp^{N-1}$,
so every pure state is $SU(N)$--coherent
and their Husimi distributions have the same 
shape and differ only by the localization in $\Omega$.

The $SU(N)$ coherent states may be defined 
according to the general group-theoretical approach by Perelomov,
by a set of generators of $SU(N)$ acting on the
distinguished reference state $|0\rangle$.
We are going to work with the degenerate
representation of $SU(N)$ only, in which
a coherent state may be parametrized by $N-1$
complex numbers $\gamma_i$,
\begin{equation}
|\alpha_N\rangle=|\gamma_1,\hdots,\gamma_{N-1}\rangle:=
e^{\gamma_1 J_1} \cdots e^{\gamma_{N-1} J_{N-1}} |0\rangle \ ,
\label{SUNCS}
\end{equation}
where operators $J_i$
may be interpreted as lowering operators \cite{GS93}.
In language of the $N$--level atom they 
couple the highest $N$-the level with the $i$-th one (see e.g. \cite{GK98}).
In the simplest case of $SU(2)$ coherent states
the reference state $|0\rangle$ is equal to the
eigenstate $|j,j\rangle$
of the angular momentum's $z$-component $J_z$
with maximal eigenvalue,
while the lowering operator reads $J_1=J_-=J_x-iJ_y$.

To characterize quantitatively the localization
of an analyzed state $\varrho$ in the phase space
we compute its Husimi distribution $H_{\varrho}$ 
with respect to $SU(N)$--CS
and analyze the moments $m_q$ of the distribution

\be
m_q(\varrho)=\int_{\Omega_N}
d\mu(\alpha)\hspace{.2cm}\left(H_{\varrho}(\alpha)\right)^q \ .
\label{mom}
\ee
 Here $d\mu(\alpha)$ denotes the unique, unitarily invariant measure 
 on $\cp^{N-1}$, also called Fubini--Study measure.
The measure $d\mu$ is normalized such that for any state $\varrho$
the first moment $m_1$ is equal to unity, so the non--negative
Husimi distribution $H_{\varrho}$ may be regarded as a
phase-space probability distribution.

The definition of the moments $m_q$ is not restricted to 
integer values of $q$.
However, from a practical point of view it will be easier to perform
the integration for integer values of $q$.
But once $m_q$ are known for all integer $q$,
there is a unique analytic extension to complex (and therefore also real) q,
as integers are dense at infinity.\\
Another quantity of interest is the Wehrl entropy $S_W$, 
defined as \cite{We78}

\be
S_W(\varrho)=-\int_{\Omega_N}d\mu (\alpha)\hspace{.2cm}H_{\varrho}
(\alpha) \ln H_{\varrho}(\alpha) \  .
\ee
Again, performing the integration might be difficult.
But if all the moments $m_q$ are known in the vicinity of $q=1$,
one can  derive $S_W$ quite easily.
Using 
$\partial H^q /\partial q=H^q \ln H$, one gets
\be
S_W(\varrho)=-\lim_{q\to1}\abl{m_q(\varrho)}{q}\ .
\label{limit}
\ee
The moments $m_q$ of the Husimi function allow us to write the
R{\'e}nyi--Wehrl entropy 
\be
S_{W,q}(\varrho):=
\frac{1}{1-q}\ln m_q(\varrho)\ ,
\label{REWE}
\ee
which tends to the Wehrl entropy $S_W$ for $q\to 1$.
As a Husimi function is related to a selected classical phase space, the
Wehrl entropy is also called {\it classical entropy}
\cite{We78,We79}, in contrast to the {\it von Neumann entropy}
$S_N=-\tr\varrho\ln\varrho$,
that has no immediate relation to classical mechanics.

In section \ref{monopartite} we consider how strongly a given state is
mixed and in section \ref{bipartite} we discuss the non-locality of
bipartite pure states.
For this purpose we pursue an approach inspired by
the Lieb conjecture \cite{Li78},
according to which the Wehrl entropy $S_W$ of a quantum state
is minimal if and only if $\varrho$ is coherent.
In order to measure a degree of mixing of a
monopartite  mixed state of size $N$
we therefore use the $SU(N)$ coherent state, while
for the study of entanglement of a bipartite  
pure $N \times N$ state
we use the coherent states related to the
product group $SU(N)\times SU(N)$.

\section{Monopartite systems: mixed states}
\label{monopartite}

In this paragraph we will focus on mixed states 
$\varrho\in{\mathbbm C}^{N\times N}$
acting on an $N$--dimensional  Hilbert space ${\cal H}_N$.
Such a state may appear as a reduced density matrix $\varrho_r$,
defined by the partial trace
$\varrho_r(\Psi)=\mbox{Tr}_A\ket{\Psi}\bra{\Psi}$
of a pure state
$\ket{\Psi}\in{\cal H}_N\otimes{\cal H}_N$
of a bipartite system.
One could also consider a system coupled to an environment,
in which a mixed state is obtained by tracing over the environmental
degrees of freedom.
The von Neumann entropy $S_N$ quantifies, on one hand,
the degree of mixing of $\varrho_r$,
and on the other, the nonlocal
properties of the bipartite state $\ket{\Psi}$.

To obtain an alternative measure of mixing of $\varrho$
we are going to investigate its Husimi function
defined with respect to the $SU(N)$--CS as  
$H^{(1)}_{\varrho}(\alpha)=\matel{\alpha_N}{\varrho}{\alpha_N}$.
The moments $m_q$ of the Husimi distribution 
can be expressed as functions of the eigenvalues
$\lambda_i$ of $\varrho$,
which in the case of a reduced state of a bipartite system coincide
with the Schmidt coefficients of the pure state $\ket{\Psi}$.
The derivation of the explicit result in terms of the
Euler Gamma function 
\be\ba{ll}
\ds m_q=\frac{N!\Gamma(q+1)}{\Gamma(q+N)}\mu_{q,N}
\hspace{1cm}\mbox{with}\vspace{.5cm}\\
\ds \mu_{q,N}=\sum_{i=1}^{N}\frac{\lambda_i^{q+N-1}}
{\ds\prod_{j=1,j\neq i}^N(\lambda_i-\lambda_j)}
\label{mq1}
\ea\ee
is provided in Appendix \ref{appmoments}.
Performing the limit (\ref{limit}) we find that the Wehrl entropy
$S_W$ equals the subentropy $Q(\varrho)$
up to an additive constant $C_N$,
\be
S_W(\varrho)=Q(\varrho)+C_N \ .
\label{SQPSI}
\ee
The $N$-dependent constant  can be expressed as 
\be
C_N=\Psi(N+1)-\Psi(2)=\sum_{k=2}^N 1/k \ ,
\label{CN}
\ee
where the digamma function is defined by 
$\Psi(x)=\frac{\partial\ln\Gamma(x)}{\partial x}$.
The subentropy
\be
Q(\varrho)=-
\sum_{i=1}^N\frac{\lambda_i^N\log\lambda_i}
{\ds\prod_{j=1,j\neq i}^N(\lambda_i-\lambda_j)}
=: Q(\vec{\lambda})
\label{subentr}
\ee
was defined in an information theoretical context \cite{JRW94}.
It is related to the von Neumann entropy $S_N(\varrho)$
in the sense that both quantities
give the lower and the upper bounds for the
information which may be extracted from the state $\varrho$ \cite{JRW94}.
The subentropy $Q(\varrho)$ takes its minimal value $0$ for pure states
with only one non vanishing eigenvalue.
Therefore we define the entropy excess $\Delta S$ as
\be
\Delta S(\varrho) = S_W(\varrho)-S_W(\rho_\psi)\ ,
\ee
where $\rho_\psi$ refers to an arbitrary pure state
and the Wehrl entropy of an arbitrary $N$-dimensional
pure state is given by $S_W(\rho_{\psi})=C_N$.
Hence the entropy excess, equal to the subentropy
$\Delta S(\varrho)=Q(\varrho)$,
is non negative and equal to zero only for pure states.

As a byproduct we find a bound on $S_W(\varrho)$.
The subentropy $Q(\varrho)$ is known not to be larger than the von
Neumann entropy $S_N$ \cite{JRW94}.
Using this fact and (\ref{SQPSI}) we find that
$S_W(\varrho) \le S_N(\varrho)+C_N$.
As it is also known that the von Neumann entropy $S_N$ is not larger
than the Wehrl entropy $S_W(\varrho)$ \cite{We78},
we end up with the following upper and lower bound on $S_W(\varrho)$
\be
S_N(\varrho)+C_N \ge S_W(\varrho)\ge S_N(\varrho)\ .
\ee

\section{Bipartite systems: pure states}
\label{bipartite}

Let us now focus on bipartite systems described by a Hilbert space ${\cal H}$
that can be decomposed into a tensor product
${\cal H}={\cal H}_A\otimes{\cal H}_B$
of two sub-spaces.
With a local unitary transformation
${\cal U}_l={\cal U}_A\otimes{\cal U}_B$ any pure state
$\ket{\Psi}\in{\cal H}$ can be transformed to 
its {\it Schmidt form} \cite{Pe95}
\begin{equation}
\ket{\Psi}=\sum_{i=1}^N\sqrt{\lambda_i}\ket{i}_A\otimes\ket{i}_B \ ,
\label{Scmid}
\end{equation}
where $N=$min(dim ${\cal H}_A$,dim ${\cal H}_B$)
and the real prefactors $\lambda_i$ called {\it Schmidt coefficients}
are the eigenvalues of the reduced density matrix $\varrho_r$.
Thanks to the {\it Schmidt decomposition} we can assume
$\mbox{dim}\hspace{.1cm}{\cal H}_A=\mbox{dim}\hspace{.1cm}{\cal H}_B=N$
without loss of generality.
Following an idea of Sugita \cite{Su03} we consider the question
whether the Wehrl entropy of a bipartite state can serve as a measure of entanglement.
He defined  a coherent state $\ket{\alpha_2^{(M)}}$ of an $M$-partite qubit
system as the tensor product of $M$ coherent states $\ket{\alpha_2^{(1)}}$
of single qubit systems.
We generalize this approach for bipartite systems, omitting the
restriction to qubits and calculate all moments and the Wehrl
entropy of the respective Husimi functions.
For a bipartite pure state $\ket{\Psi}\in{\cal H}$,
we use the tensor product of two SU$(N)$-coherent
states
$\ket{\alpha_N^{(2)}}=\ket{\alpha_N}^{\otimes 2}\in{\cal H}$
to define a Husimi function.
More explicitly one can write
$\ket{\alpha_N^{(2)}}=\ket{\alpha_N}_A\otimes\ket{\alpha_N}_B$,
with $\ket{\alpha_N}_A\in{\cal H}_A$ and $\ket{\alpha_N}_B\in{\cal H}_B$
(we will drop the index referring to the subsystem,
 wherever there is no ambiguity).
Such bipartite coherent states were used already in the original
paper of Wehrl \cite{We79}.

The Husimi function $H$ of a bipartite state $\ket{\Psi}$ is then given by

\be
H_{\Psi} (\alpha_A, \alpha_B) =
\bigl| \bra{\Psi}(\ket{\alpha_N}_A \otimes \ket{\alpha_N}_B)\bigr|^2\ .
\label{Husim2}
\ee

By definition any product state is a $SU(N)\times SU(N)$ coherent state.
According to the Lieb conjecture, originally formulated
for $SU(2)$ coherent states, the Wehrl entropy is minimal for coherent states
\cite{Li78}.
Hence it is natural to expect that the Wehrl entropy
\begin{eqnarray}
S_W(\psi)=-\int_{\Omega_N}d\mu_A(\alpha)
\int_{\Omega_N}d\mu_B(\alpha)\vspace{.5cm}\nonumber\\
H_\Psi(\alpha_A,\alpha_B) \ln H_\Psi(\alpha_A,\alpha_B)
\end{eqnarray}
provides a measure of how ``incoherent'' a given state is.
Due to the equivalence of ``coherence'' and separability we expect that
$S_W$ can also serve as an entanglement measure.
In the following we discuss properties of $S_W$ and show
that it satisfies all requirements of entanglement monotones \cite{VP98,Vi00}.

There is a one-to-one correspondence between states and Husimi functions.
Concerning entanglement,
two different states that are connected by a local unitary
transformation are considered equivalent.
As by construction the moments $m_q$ of $H$ are invariant
under local unitary transformations,
they reflect this equivalence and therefore can be expected to be good
quantities to characterize the nonlocal properties of a bipartite state.
For a pure state $\ket{\Psi}\in{\cal H}_A \otimes {\cal H}_B$,
there are $N-1$ independent moments,
determining $N-1$ independent Schmidt coefficients
(one coefficient is determined by the normalization).
In our case the moments read
\be
m_q=
\int_{\Omega_N}\hspace{-.3cm}d\mu_A(\alpha_A)
\int_{\Omega_N}\hspace{-.3cm}d\mu_B(\alpha_B)
\hspace{.2cm}
\Bigl[ H_{\Psi}(\alpha_A,\alpha_B) \Bigr]^q , 
\label{mq2}
\ee
where the integration is performed over 
the Cartesian product $\Omega_N \times \Omega_N$,
being the space of all product pure states
of the bipartite system.
The proper normalization of $d\mu_A$ and $d\mu_B$ assures that
$m_1=1$.
The required $N-1$ monotones can be provided by $m_q$ with $q=2,\hdots,N$.
The moments can be expressed as a function of the Schmidt coefficients
$\lambda_i$, defined in eq (\ref{Scmid})
\be
m_q=
\left(\frac{N!\Gamma(q+1)}{\Gamma(q+N)}\right)^2\mu_{q,N}
\label{moments}\ee
and are related to the monopartite moments (eq. \ref{mq1})
by a multiplicative factor.
The derivation of this result is provided in Appendix \ref{appmoments}.
Having closed expressions for the moments that can be extended
to real $q$, one easily finds the bipartite Wehrl entropy
\begin{eqnarray}
S_W(\psi)=-\int_{\Omega_N}d\mu_1(\alpha_A)
\int_{\Omega_N}d\mu_2(\alpha_B)\nonumber\\
H_\Psi(\alpha_A,\alpha_B)\ln H_\Psi(\alpha_A,\alpha_B)\ .
\end{eqnarray}
Performing the limit (\ref{limit}) one finds that $S_W$ equals
the corresponding monopartite quantity up to an additive constant
\be
S_W(\Psi)=Q(\vec{\lambda})+2C_N\ .
\ee

\begin{figure*}
  \epsfig{file=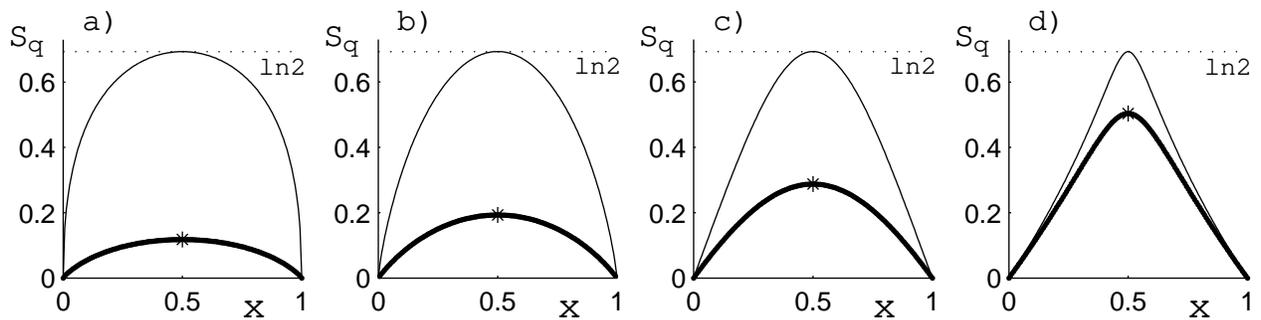}
  \caption{
    R\'enyi entropy $S_q$ (thin line) and R\'enyi subentropy $Q_q$ (thick line)
    line for the $N=2$ probability vectors as
    a function of the one independent variable $x$ for 
    $q=1/2$ (a), $q=1$ (b), $q=2$ (c) and $q=10$ (d).
    The maximum of the R\'enyi subentropy is increasing with $q$ whereas
    the maximum of the R\'enyi entropy is independent of $q$.
    In the limit $q\to\infty$ both quantities converge to
    $-\ln\lambda_{max}$, where $\lambda_{max}$ is the largest
    component of the vector $\vec \lambda$.}
  \label{rensub2}
\end{figure*}

The subentropy 
$Q(\lambda)$ takes its minimal value $0$
if and only if a given state is separable,
{\it i.e.} if all but one Schmidt coefficients vanish.
Therefore we define the bipartite entropy excess $\Delta S_W(\Psi)$
\be
\Delta S_W(\Psi) = S_W(\Psi)-S_W(\phi_{sep}) = Q(\varrho_r)
\ee
that is vanishing if and only if $\ket{\Psi}$ is separable.
Furthermore, one can also consider the R\'enyi--Wehrl entropy
$S_{RW}^{(q)}=\frac{1}{1-q}\ln m_q$.

In order to use the moments $m_q$ as entanglement monotones,
it  is conventional to rescale them such that they vanish for
separable states and are positive for entangled ones
\be
\ds M_q=
\frac{1}{1-q}(\mu_{q,N}-1)\ .
\label{rescale}
\ee
These quantities are analogous to the Havrda-Charvat entropy
(also called Tsallis entropy) \cite{HC67,Ts88}
and in the limit $q\to 1$ one obtains the subentropy
\be
\lim_{q\to 1}M_q=Q(\vec{\lambda})\ .
\ee

\section{R\'enyi subentropy}

As discussed in previous sections,
the excess of the Wehrl entropy $\Delta S$ may be used
as a measure of the degree of mixing for monopartie states,
or degree of entanglement for pure states of bipartite systems.
Since the moments of the Husimi distribution are found, 
we may extend the above analyzis for the 
R{\'e}nyi--Wehrl entropy $S_{W,q}$ defined by (\ref{REWE}).

Considering, for instance, the case of mixed states of
a monopartite system we use (\ref{mq1})
to find the minimal R{\'e}nyi--Wehrl entropy 
attained for pure states $\varrho_\psi$, 
\be
C_{N,q}=S_{W,q}(\varrho_\psi)=
\frac{1}{1-q} \ln \frac{N! \Gamma(q+1)}{\Gamma(q+N)}\ ,
\label{CNq}
\ee
which for $q\to 1$ reduces to (\ref{CN}).
In an analogy to (\ref{SQPSI})
we define the R{\'e}nyi--Wehrl entropy excess 
\be
\Delta S_{W,q}(\varrho)=S_{W,q}(\varrho)-C_{N,q}\ .
\label{SQPSIq}
\ee
Applying (\ref{mq1}) it is straightforward
to obtain result in terms of the eigenvalues 
$\lambda_i$ of the analyzed state $\varrho$ 
\be
\Delta S_{W,q}(\varrho)=
\frac{1}{1-q}\ln
\sum_{i=1}^{N}\frac{\lambda_i^{q+N-1}}
{\ds\prod_{j=1,j\neq i}^N(\lambda_i-\lambda_j)}
=:Q_q({\vec \lambda})\ .
\label{Qq}
\ee
This result shows that the excess of the R{\'e}nyi Wehrl entropy
$Q_q(\varrho)=Q_q({\vec \lambda})$ 
may be called {\sl R{\'e}nyi subentropy}
since for $q\to 1$ it tends to the subentropy 
(\ref{subentr}).
On one hand it may be treated as a function  
of an arbitrary quantum state $\varrho$,
on the other it may be defined for an 
arbitrary classical probability vector $\vec \lambda$.  

\begin{figure*}
  \epsfig{file=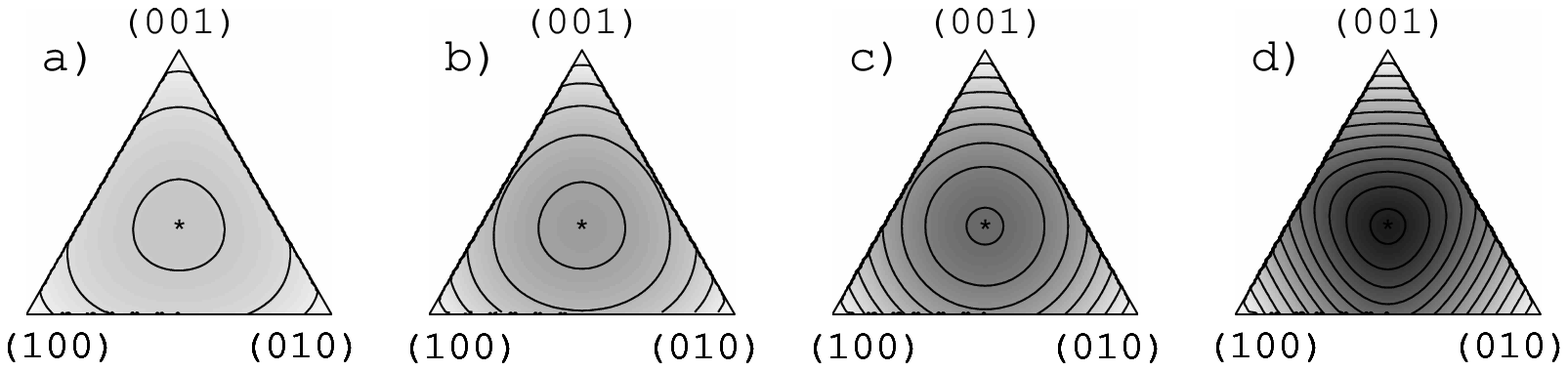}
  \caption{
    R\'enyi subentropy $Q_q$ as a function of two independent 
    variables of the $N=3$ probability distributions,
    $q=1/2$ (a), $q=1$ (b), $q=2$ (c) and $q=5$ (d).
    Gray scale is used to represent the values of subentropy:
    the higher $q$ the larger the maximum of $Q_q$
    at the center ${\vec \lambda_*}=(1/3,1/3,1/3)$,
    which refers to maximally entangled states.} 
  \label{rensub3}
  \epsfig{file=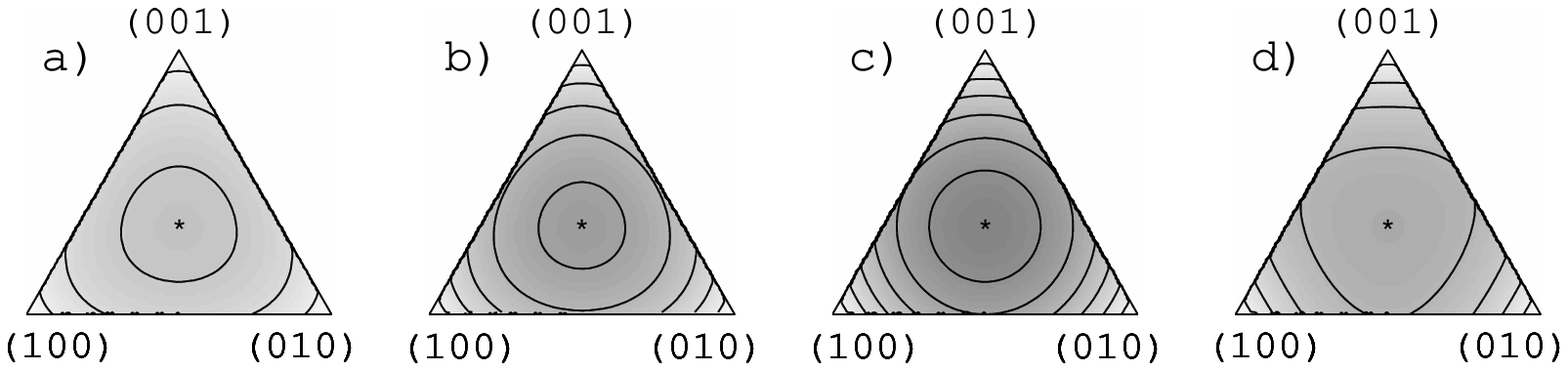}
  \caption{
    Rescaled moments $M_q$
    as a function of two independent variables for
    $N=3$ probability vectors, 
    $q=1/2$ (a), $q=1$ (b), $q=2$ (c) and $q=5$ (d).
    Labels at corners identify pure separable states.
    Note that in general $M_q$ is not a monotonically increasing
    function of q.}
  \label{renmom3}
\end{figure*}
In a sense $Q_q$ is a quantity 
analogous to the R\'enyi entropy $S_q$,
which may be defined for an arbitrary 
probability vector 
\be
S_q(\vec{\lambda})=\frac{1}{1-q}\ln \sum_{i=1}^N \lambda_i^q .
\ee
In the case of quantum states $S_q$ is defined as a
 function of the $q$-th moment of the respective state
\be
S_q(\varrho)=\frac{1}{1-q}\ln(\mbox{Tr}\varrho^q) . 
\ee

In the following we will sketch some properties of $Q_q(\vec{\lambda})$ 
as a function of a classical propability vector.
All the considerations are also applicable in the quantum case,
particularly if we use $Q_q$ as an entanglement monotone 
using the moments discussed in the preceding section.
Let us define two distinguished propability vectors
which correspond to extreme cases:
$\vec{\lambda}_\ast$ with $\lambda_i^{\ast}=\frac{1}{N}$ ($i=1,\ldots,N$)
describes the maximal random event,
whereas $\vec{\lambda}_s$ with $\lambda_i^{(s)}=\delta_{i,j}$ with $j\in [1,\ldots,N]$ describes an event with a certain result.
For a bipartite quantum system a vector of Schmidt coefficients given by
$\vec{\lambda}_\ast$ represents a maximally entangled state, whereas
$\vec{\lambda}_s$ describes a separable state.

The R\'enyi subentropy $Q_q({\vec \lambda})$ has the following properties
\begin{itemize}
\item[i.)]
  $Q_q({\vec \lambda}_s)=0$ for any $q$.
\item[ii.)]
$Q_q$ takes its maximal value
$Q_q^{max}=\frac{1}{1-q}\ln \frac{\Gamma(q+N)}{\Gamma(q+1)N!}\frac{1}{N^q}$
for $\vec{\lambda}_\ast$.
For $q\to 1$ one has $Q_q^{max}=C_N-\ln N$.
\item[iii.)]
As $m_0^{(1)}=1$ one immediately has
$\lim_{q\to 0}Q_q(\vec\lambda)=0$
for any vector $\lambda$.
\item[iv.)]
For $q\to 1$ one obtains the regular subentropy, 
 $\lim_{q\to 1}Q_q ({\vec \lambda})=Q({\vec \lambda})$. 
\item[v.)]
For $q\to\infty$ one gets
$\lim_{q\to\infty}Q_q({\vec \lambda})
=-\ln \lambda_{max}$, where $\lambda_{max}$ is the
largest entry of $\lambda$.
Hence this limit coincides with the limit $q\to \infty$
 of the R\'enyi entropy $S_q(\vec{\lambda})$.
\item[vi.)]
$Q_q$ is {\sl expansibile}, i.e. it does not vary if the 
probability vector $\vec\lambda$ is extended by zero,
$Q_q(\lambda_1,\dots,\lambda_N)=Q_q(\lambda_1,\dots,\lambda_N,0)$.
\end{itemize}

A discrete interpolation between the subentropy $Q$ and the Shannon entropy $S$
was recently proposed in  \cite{WOT03}.
Note that a continuous interpolation 
between these quantities may be obtained 
by the R{\'e}nyi subentropy $Q_q$ for $q$ increasing from unity to 
infinity combined with the R{\'e}nyi entropy 
$S_q$ for $q$ decreasing from infinity to unity.
It is well known \cite{BS93}
that the R\'enyi entropy $S_q$ is both a non-increasing 
function of $q$ and convex with respect to $q$.

\begin{figure}
\epsfig{file=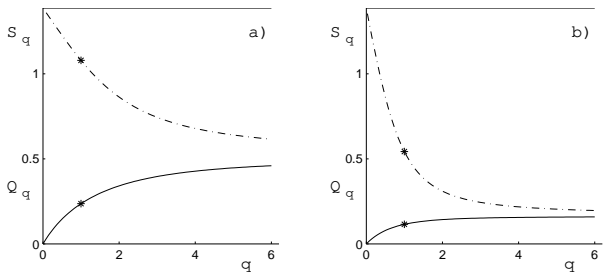,width=8cm}
\caption{\label{subeq4}
R\'enyi entropy $S_q$ (dashed lines) and
R\'enyi subentropy $Q_q$ (solid lines) as a 
function of $q$ for two exemplary probability distributions.
We have chosen $N=4$ probability vectors with 
power law distributed components, 
$p_j\propto j^{\kappa}$ with $\kappa=3/2$ in a) and $\kappa=3$ in b).
Stars at $q=1$ represent Shannon entropy (upper) and subentropy (lower).
We conjecture that $Q_q$ is both non-decreasing and concave with
respect to the R\'enyi parameter $q$.}
\end{figure}

We conjecture that the R\'enyi subentropy has opposite properties:
it is non--decreasing as a function of $q$
and it is concave with respect to $q$.
See Fig. \ref{subeq4} for some exemplary cases.
Fig. \ref{rensub2} presents both quantities for $N=2$,
while Fig. \ref{rensub3} and Fig. \ref{renmom3} present the curves of
isoentropy $Q_q$
and equal moments $M_q$ obtained for $N=3$.

\section{Schur concavity}

So far we defined rescaled moments, R\'enyi entropy and R\'enyi subentropy.
All these quantities are constructed in such a way that they are vanishing exactly for separable states.
In order to legitimately use these quantities as entanglement monotones,
we still have to show, that they are non increasing under local
operations and classical communication,
or analogously that they are Schur concave \cite{Ni99,Vi00},
{\it i.e.}
\be
\vec{\lambda}\prec\vec{\xi}
\hspace{.5cm}\Rightarrow\hspace{.5cm}
\Delta S_W(\vec{\lambda})\ge \Delta S_W(\vec{\xi})\ ,
\label{schur}\ee
or equivalently for the other considered quantities.
The expression $\vec{\lambda}\prec\vec{\xi}$
means that $\vec{\lambda}$ is majorized by $\vec{\xi}$,
{\it i.e.} the components of both vectors listed in increasing order
satisfy $\sum_{i=1}^{j}\lambda_i\le\sum_{i=1}^{j}\xi_i$
for $0<j\le N$.
In order to be Schur concave,
$\Delta S_W(\vec{\lambda})$ has first to be invariant under
exchange of any two arguments, which is obviously the case,
and second it has to satisfy \cite{An89}
\be
(\lambda_1-\lambda_2)
\left(
\frac{\partial\Delta S_W}{\partial\lambda_1}-
\frac{\partial\Delta S_W}{\partial\lambda_2}\right)\le 0\ .
\label{inequ}
\ee
For convenience we will first consider $\mu_{q,N}$.
Once we know about Schur convexity or Schur concavity of $\mu_{q,N}$
one can easily deduce Schur concavity of all
the other discussed quantities.
The quantities $\mu_{q,N}$ can be expressed by the following representation
(see eq. \ref{intexp})
$\mu_{q,N}=
\frac{\Gamma(q+N)}{\Gamma(q+1)}\int_\Delta dx(\vec\lambda\vec x)^q
$,
where $dx$ is a short hand notation for $dx_1\hdots dx_N$ and the
integration is performed over the simplex $\Delta$
containing all propability vectors $\vec x$ with $\sum_ix_i=1$.
Making use of this relation and assuming $\lambda_1>\lambda_2$
without loss of generality, we obtain
\be
\frac{\partial\mu_{q,N}}{\partial\lambda_1}-
\frac{\partial\mu_{q,N}}{\partial\lambda_2}=
\frac{\Gamma(q+N)}{\Gamma(q+1)}\int_\Delta dx\
(\vec\lambda\vec x)^{q-1}(x_1-x_2)\ .
\ee
The full measure of the propability simplex $\Delta$ can be divided into two symmetric parts,
the first part specified by $x_1<x_2$, the second one by
$x_1>x_2$.
After exchanging the labels $x_1$ and $x_2$ in the second part,
the first and the second part coincide and one gets
\begin{eqnarray}
\frac{\partial\mu_{q,N}}{\partial\lambda_1}-
\frac{\partial\mu_{q,N}}{\partial\lambda_2}=
\frac{\Gamma(q+N)}{\Gamma(q+1)}\int_{\Delta_>}dx
(x_1-x_2)(\lambda_1-\lambda_2)
\nonumber\\
\left(\hspace{-.1cm}\Bigl(\hspace{-.1cm}
\lambda_1x_1\hspace{-.05cm}+\hspace{-.05cm}\lambda_2x_2\hspace{-.05cm}+\hspace{-.1cm}
\sum_{i>2}\hspace{-.1cm}\lambda_ix_i\Bigr)^{q-1}\hspace{-.4cm}-
\Bigl(\lambda_1x_2\hspace{-.05cm}+\hspace{-.05cm}\lambda_2x_1\hspace{-.05cm}+\hspace{-.1cm}
\sum_{i>2}\hspace{-.1cm}\lambda_ix_i\Bigr)^{q-1}\hspace{-.1cm}\right)\ .
\end{eqnarray}
Since $x_1>x_2$ and $\lambda_1>\lambda_2$, one has
$x_1\lambda_1+x_2\lambda_2>x_2\lambda_1+x_1\lambda_2$.
Therfore for $q>1$ the integrand is non-negative and so is the integral.
For $0<q<1$ the integrand is non-positive.
Thus $\mu_{q,N}$ is Schur convex for $q>1$ and Schur concave for $0<q<1$.
Using this, one easily concludes that
\be
(\lambda_1-\lambda_2)
\left.\left(
\frac{\partial}{\partial\lambda_1}-\frac{\partial}{\partial\lambda_2}\right)
\frac{\partial \mu_{q,N}}{\partial q}
\right|_{q=1}\ge 0\ .
\ee
As $\Delta S_W$ can be expressed as
$\Delta S_W=-\lim_{q\to 1}\frac{\partial \mu_{q,N}}{\partial q}$,
Schur concavity of the subentropy can directly inferred from the
corresponding properties of $\mu_{q,N}$.
Also the rescaled moments (\ref{rescale}) are Schur concave for all
values of $q$, as $1-q$ is negative for $q>1$ where $\mu_{q,N}$
is Schur convex.

For the R\'enyi subentropy $Q_q$ with $q\ne 1$ we have
\begin{eqnarray}
(\lambda_1-\lambda_2)
\left(\frac{\partial Q_q}{\partial\lambda_1}-
 \frac{\partial Q_q}{\partial\lambda_2}\right)=\nonumber\\
\frac{1}{1-q}\frac{1}{Q_q}(\lambda_1-\lambda_2)
\left(\frac{\partial \mu_{q,N}}{\partial\lambda_1}-
 \frac{\partial \mu_{q,N}}{\partial\lambda_2}\right)\ .
\end{eqnarray}
$Q_q$ is a positive quantity.
Using Schur concavity of $\mu_{q,N}$ for $0<q<1$ and  Schur convexity
for $q>1$,
we conclude that $Q_q$ is Schur concave for all positive values of $q$.

Since we have shown that the subentropy, the rescaled moments and the
R\'enyi subentropy vanish if and only if the
considered state is separable and that these quantities 
are Schur concave, 
these quantities are entanglement monotones
and may serve as legitimate measures of quantum entanglement.
Let us emphasize that the monotones found in this work
differ from the R{\'e}nyi entropies \cite{ZB02} and  
the elementary symmetric polynomials of the Schmidt 
coefficients \cite{SZK02} 
and cannot be represented as a functions 
of one of these quantities.

\section{Outlook}

Defining the Husimi function of a given  
pure state of a bi-partite $N\times N$ 
system with respect
to coherent states related to $SU(N)\times SU(N)$ product groups 
and computing the Wehrl entropy 
allows us to establish a link between the 
phase space approach to quantum mechanics
and the theory of quantum entanglement.

This approach may be considered as an example
of a more general method of measuring 
the closeness of an analyzed state
to a family of distinguished states.
This idea, inspired by the work of Sugita \cite{Su03},
may be outlined in the following set-up:

\smallskip \noindent
{\bf a)} Consider a set of (pure or mixed) states you wish to analyze

\smallskip \noindent
{\bf b)}
Select a family of distinguished (coherent) states $\ket{\alpha}$,
   related to a given symmetry group and
   parametrized by a point on a certain manifold $\Omega$.
   This family of states has to satisfy the identity resolution,
   $\int_{\Omega} |\alpha\rangle \langle \alpha |
    {\rm d} \Omega ={\mathbb I}$.

\smallskip \noindent
{\bf c)} Define the Husimi distribution with respect to the
   'coherent' states and find the Wehrl entropy for
    a coherent state, $S_{min}=S(|\alpha\rangle)$

\smallskip \noindent
{\bf d)} Calculate the Wehrl entropy for the analyzed state,
   $S(|\psi\rangle)$

\smallskip \noindent
{\bf e)} Compute the {\sl entropy excess}
   $\Delta S=S(|\psi\rangle)-S_{min}$,
   which characterizes quantitatively to what extend
   the analyzed state $|\psi\rangle$ is not 'coherent'.

For instance, analyzing the space of pure states
of size $N$
we may distinguish the spin $SU(2)$ coherent states,
or, in general the $SU(K)$ coherent states
(with $K<N$), parametrized by a point on
${\mathbb C}P^1$ and on  ${\mathbb C}P^{K-1}$,
respectively.  In the case of $N\times L$
composite quantum system we may distinguish
the $SU(N)\otimes SU(L)$ coherent states.
They form the set of product (separable)
states and are labeled by a point on
${\mathbb C}P^{N-1} \times {\mathbb C}P^{L-1}$.
For mixed states of size $N$
we may select $SU(N)$ coherent states, i.e.
all pure states. In all these three
cases we define the same quantity,
entropy excess
$\Delta S$, which has
entirely different physical meaning.
It quantifies the degree of non--$SU(K)$--coherence,
the degree of entanglement and the
degree of mixing, respectively,
as listed in Table \ref{tab1}.

\begin{table*}
\begin{tabular}[t]{|c||c||c|c|c|c|}\hline
States & Mixed & Pure  & Pure & Pure & Pure \\
\hline
Systems & simple & simple & simple & bipartite &multipartite \\
\hline
\parbox{2.2cm}{\centering Coherent states} &
$SU(N)$ CS & $SU(2)$ CS & $SU(K)$ CS &
$SU(N)\otimes SU(N)$ CS & $SU(N)^{\otimes M}$ CS \\
&
$|\alpha_{N}\rangle\in {\mathbb C}P^{N-1}$ &
$|\alpha_{2}\rangle\in {\mathbb C}P^{1}$ &
$|\alpha_{K}\rangle\in {\mathbb C}P^{K-1}$ &
$|\alpha_{N}\rangle^{\otimes 2}\in {\mathbb C}P^{N^{2}-1}$ &
$|\alpha_{N}\rangle^{\otimes M}\in {\mathbb C}P^{N^{M}\!-\! 1}$ \\
\hline
\parbox{2.2cm}{\centering Minimal Wehrl entropy\vspace{.1cm}} &
$S_{\min}=S(|\alpha_{N}\rangle)$ &
$S_{\min}=S(|\alpha_{2}\rangle)$ &
$S_{\min}=S(|\alpha_{K}\rangle)$ &
$S_{\min}=2S(|\alpha_{N}\rangle)$ &
$S_{\min}=MS(|\alpha_{N}\rangle)$ \\
\hline
Generic states & $\rho:{\cal H}_N\rightarrow {\cal H}_N$ &
$|\psi\rangle\in {\mathbb C}P^{N-1}$ &
$|\psi\rangle\in {\mathbb C}P^{N-1}$ &
$|\Psi\rangle\in ({\mathbb C}P^{N-1})^{\times 2}$ &
$|\Psi\rangle\in ({\mathbb C}P^{N-1})^{\times M}$ \\
\hline
\parbox{2.2cm}{\centering Wehrl entropy excess $\Delta S$\vspace{.1cm}} &
$S(\rho)-C_N$ & $S(|\psi\rangle)-C_2$ &
$S(|\psi\rangle)-C_K$ & $S(|\Psi\rangle)-2C_N$ &
$S(|\Psi\rangle)-MC_N$ \\
\hline
\parbox{2.2cm}{measures degree of} &
mixing &
\parbox{2.1cm}{\centering non $SU(2)$ coherence\vspace{.1cm}} &
\parbox{2.3cm}{\centering non $SU(K)$ coherence\vspace{.1cm}} &
\parbox{2.5cm}{\centering bi-partite entanglement\vspace{.1cm}} &
\parbox{2.5cm}{\centering non-$M$-partite separability\vspace{.1cm}}\\
\hline
\end{tabular}
\caption{\label{tab1} Wehrl entropy excess in various set-ups. This
quantity measures the `distance' of a given mixed state
 from the subset of pure states; the `distance' of a given pure state
 from the subset of all $SU(K)$ coherent states;
 the distance of a given pure state of a composite system
 from the subset of all separable (product) states.}
\end{table*}
Describing the items a)--e) of the above procedure
we have implicitly assumed that
the Wehrl entropy is minimal if and only if
the state is 'coherent'. This important
point requires a comment, since the status
of this assumption is different in the cases
discussed.
For the space of $N$--dimensional pure states analyzed by Husimi functions
computed with respect to $SU(2)$ coherent states this
statement became famous as the Lieb conjecture \cite{Li78}.
Although it was proven in some special cases of
low dimensional systems \cite{Sc99,GZ01,Su02},
and is widely believed to be true for an arbitrary
dimensions, this conjecture still awaits
a formal proof. On the other hand,
it is easy to see that
the Wehrl entropy of a mixed state is minimal
if and only if the state is pure,
or the Wehrl entropy of a bipartite pure state
is minimal if and only if the state is separable.
This is due to the fact that in both cases
the entropy excess is equal to the subentropy
which is non-negative and is equal to zero
only if the state is pure \cite{JRW94}.

One of the main advantages of our approach is that it can easily be generalized to the problem
of pure states of multipartite systems
-- see the last column of Table 1.
It is clear that the entropy excess of $|\Psi\rangle$ 
is equal to zero if $|\Psi\rangle$ is a product state,
but the reverse statement requires a 
formal proof. Moreover, the Schmidt decomposition 
does not work for a three (or many)--partite case.
Thus in order to compute explicitly the 
entropy excess in these cases,
one has to consider by far more terms than in the bipartite case.
In the special case of three qubits any pure state may be characterized by a
set of five parameters \cite{AAJ01,Sud01},
and it would be interesting to express the
entropy excess of an arbitrary three--qubit
pure states as a function of these parameters.
The second moment for this system was already calculated \cite{Su03}
but expressions for other moments or for the Wehrl entropy are still
missing.

\section{Acknowledgment}
We are indebted to Andreas Buchleitner, Marek Ku\'s,
Bernard Lavenda and Prot Pako\'nski for fruitful discussions, comments
and remarks
and acknowledge fruitful correspondence with Ayumu Sugita.
Financial support by Volkswagen Stiftung and a research grant by
Komitet Bada{\'n} Narodowych is gratefully acknowledged.

\appendix{}
\section{Moments of the Husimi function}
\label{appmoments}

As already mentioned, every pure state belonging to an $N$--dimensional 
Hilbert space ${\cal H}_N$ is a $SU(N)$ coherent state and vice versa.
Therefore, following (\ref{SUNCS}), 
 we can parametrize all $SU(N)$ coherent states by 
\be
\ket{\alpha}=
 \Bigl( 1-\sum_{i=1}^{N-1} x_i \Bigr)^{1/2} \ket{0}+
\sum_{i=1}^{N-1}\sqrt{x_i}e^{i\varphi_i}\ket{i}\ ,
\label{coher}
\ee
with
$0\le x_i\le 1-\sum_{j=i+1}^{N-1}x_j$ and
$d\mu=\frac{N!}{(2\pi)^{N-1}}\prod_{i=1}^{N-1}dx_i \hspace{.1cm}d\varphi_i$.
The states $\ket{i}$,
${i=0,\ldots,N-1}$ form an orthonormal basis, while 
$\ket{0}$ is the reference state.
The Husimi function $H_{\Psi}$ of a pure state $\ket{\Psi}$
with respect to $SU(N)\otimes SU(N)$-coherent states
$\ket{\alpha_N^{(2)}}=\ket{\alpha_N}_A\otimes\ket{\alpha_N}_B$
reads
\begin{eqnarray}
H_{\Psi}=\sum_{\nu,\mu=1}^N\sqrt{\lambda_{\nu}\lambda_{\mu}}
_A\ol{\alpha_N}{\nu}_A\hspace{.1cm}
_A\ol{\mu}{\alpha_N}_A\hspace{.1cm}\nonumber\\
_B\ol{\alpha_N}{\nu}_B\hspace{.1cm}
_B\ol{\mu}{\alpha_N}_B\ ,
\end{eqnarray}
where $\ket{\Psi}$ is represented in its Schmidt basis, 
$\ket{\Psi}=\sum_\nu\sqrt{\lambda_\nu}\ket{\nu}_A\otimes\ket{\nu}_B$.
The $q$-th moment is than given by
\begin{eqnarray}
m_q=
\sum_{{\nu_1...\nu_q=1}\atop{\mu_1...\mu_q=1}}^N
\sqrt{\lambda_{\nu_1}...\lambda_{\nu_q}\lambda_{\mu_1}...\lambda_{\mu_q}}
\nonumber\\
f_q^2({\cal N}_1,...,{\cal N}_N,{\cal M}_1,...,{\cal M}_N)\ ,
\end{eqnarray}
with
\be
f_q=
\int d\mu(\alpha)
\ol{\alpha}{\nu_1}\ol{\mu_1}{\alpha}\ol{\alpha}{\nu_2}\ol{\mu_2}{\alpha}...
\ol{\alpha}{\nu_q}\ol{\mu_q}{\alpha}
\ee
The integers ${\cal N}_i$ and ${\cal M}_i$ count how often the states
$\ket{\nu_i}$ and $\bra{\mu_i}$
appear in $f_q$.
Note that one has to integrate separately over both sub-systems.
Due to the symmetry of the Schmidt decomposition with respect to
the two subsystems,
both integrations lead to the same result and it is just the square of
the integral over $f_q$ that is entering $m_q$.
Using the parametrization (\ref{coher}) $f_q$ can be expressed as
\begin{eqnarray}
f_q&=&\frac{N!}{(2\pi)^{N-1}}
\prod_{i=1}^{N-1}
\int_0^{x_i^{(max)}}\hspace{-.7cm}dx_i\int_0^{2\pi}\hspace{-.3cm}d\varphi_i\hspace{.2cm}
e^{i({\cal N}_i-{\cal M}_i)\varphi_i}\nonumber\\
&&x_i^{\frac12({\cal N}_i+{\cal M}_i)}
(1-\sum_{j=1}^{N-1}x_j)^{q-\frac12\sum_{\chi=1}^{N-1}{\cal
N}_\chi+{\cal M}_\chi}\ ,
\end{eqnarray}
where the integrations over $x_i$ are performed in increasing order of
$i$ and the upper integration limit is given by
$x_i^{(max)}=1-\sum_{j=i+1}^{N-1}x_j$.
It is useful to perform the $\varphi_i$ integrations first.
They lead to $\delta_{{\cal N}_i,{\cal M}_i}$ terms,
so that one gets
\begin{eqnarray}
f_q&=&N!
\left(
\prod_{i=1}^{N-1}
\int_0^{x_i^{(max)}}\hspace{-.7cm}dx_i\hspace{.2cm}
x_i^{{\cal N}_i}\delta_{{\cal N}_i,{\cal M}_i}
\right)\nonumber\\
&&(1-\sum_{j=1}^{N-1}x_j)^{q-\sum_{\chi=1}^{N-1}{\cal N}_\chi}\ .
\end{eqnarray}
In order to perform the integrations over $x_i$
we need to define an auxiliary function
\be
g_q(\beta,N)=
\left(\prod_{i=1}^{N-1}
\int_0^{\tilde x_i^{(max)}(\beta)}\hspace{-.7cm}dx_i\hspace{.2cm}x_i^{{\cal N}_i}\right)
(\beta-\sum_{j=1}^{N-1}x_j)^{q-\sum_{\chi=1}^{N-1}{\cal N}_\chi}\ ,
\label{def}\ee
with $\tilde x_i^{(max)}(\beta)=\beta-\sum_{j=i+1}^{N-1}x_j$.
In the following we will show that for any integer $q$
\be
g_q(\beta,N)=\beta^{(q+N-1)}
\frac{(\prod_{i=1}^{N-1}{\cal N}_i!)(q-\sum_{i=1}^{N-1}{\cal N}_i)!}
{(q+N-1)!}
\label{annahme}
\ee
holds.
According to the definition (\ref{def}),
$g_q(\beta,N)$ satisfies the following relation
\be
g_q(\beta,N)=
\int_0^\beta dx_{N-1}\hspace{.2cm}x_{N-1}^{{\cal N}_{N-1}}
g_{q-{\cal N}_{N-1}}(\beta-x_{N-1},N-1)
\label{relzwei}
\ee
Making use of
\begin{eqnarray}
g_q(\beta,1)&=&
\int_0^\beta dx_1\hspace{.2cm}x_1^{{\cal N}_1}\hspace{.1cm}
(\beta-x_1)^{q-{\cal N}_1}\nonumber\\
&=&\beta^{q+1}\frac{{\cal N}_1!(q-{\cal N}_1)!}{(q+1)!}\ ,
\label{releins}
\end{eqnarray}
we immediately have (\ref{annahme}) for $N=2$
\be
g_q(\beta,2)=
\int_0^\beta
dx_1\hspace{.2cm}x_1^{{\cal N}_1}(\beta-x_1)^{q-{\cal N}_1}\ .
\ee
Assuming that (\ref{annahme}) holds true for $g_q(\beta,N-1)$ and making use of
(\ref{relzwei}) one gets
\begin{eqnarray}
g_q(\beta,N)&=
\int_0^\beta dx_{N-1}\hspace{.2cm}x_{N-1}^{{\cal N}_{N-1}}
(\beta-x_{N-1})^{q+N-2-{\cal N}_{N-1}}\nonumber\\
&\frac{(\prod_{i=1}^{N-2}{\cal N}_i!)
(q-{\cal N}_{N-1}-\sum_{i=1}^{N-2}{\cal N}_i)!}
{(q+N-2-{\cal N}_{N-1})!}\ .
\end{eqnarray}
Using further (\ref{releins}) one has
\begin{eqnarray}
g_q(\beta,N)&=&
\beta^{q+N-1}
\frac{{\cal N}_{N-1}!(q+N-2-{\cal N}_{N-1})!}{(q+N-1)!}
\nonumber\\
&&\frac{(\prod_{i=1}^{N-2}{\cal N}_i!)
(q-\sum_{i=1}^{N-1} {\cal N}_i)!}
{(q+N-2-{\cal N}_{N-1})!}\ .
\end{eqnarray}
Finally one ends up with
\be
g_q(\beta,N)=\beta^{q+N-1}
\frac{(\prod_{i=1}^{N-1} {\cal N}_i!)
(q-\sum_{i=1}^{N-1}{\cal N}_i)!}{(q+N-1)!}\ ,
\ee
{\it i.e.} (\ref{annahme}) also holds for $g_q(\beta,N)$.
Finally for $\beta=1$, we get
\be
f_q=
N!\frac{
\left(\prod_{i=1}^{N-1} ({\cal N}_i!)\right)
\left(q-\sum_{i=1}^{N-1} {\cal N}_i\right)!}{(q+N-1)!}
\prod_{j=1}^{N-1}\delta_{{\cal N}_j,{\cal M}_j}\ .
\ee
Now we have
$
m_{q} =
\sum_{\nu_1...\nu_q=1}^N
\lambda_{\nu_1}...\lambda_{\nu_q}f_q^2
$,
where the $\delta_{n_\nu,m_\nu}$-terms assure that there are only
integer powers of the $\lambda_{\nu_i}$.
Since $f_q$ depends only on the integer powers ${\cal N}_i$
of the $\lambda_{\nu_i}$,
we need to count how many terms with fixed powers occur.
Using simple combinatorics one can see that there are
$
\left(\frac{q!}{\left(\prod_{i=1}^{N-1} ({\cal N}_i)!\right)
\left(q-\sum_{i=1}^{N-1} {\cal N}_i\right)!}\right)^2
$
terms containing the expression
$\lambda_1^{{\cal N}_1}\lambda_2^{{\cal N}_2}\ldots
\lambda_{N-1}^{{\cal N}_{N-1}}\lambda_N^{q-\sum_{i=1}^{N-1} {\cal N}_i}$.
Thus finally we get
\begin{eqnarray}
m_q&=&
\left(\frac{N!\Gamma(q+1)}{\Gamma(q+N)}\right)^2
\sum_{{\cal N}_1=0}^q
\sum_{{\cal N}_2=0}^{q-\nu_1}...
\sum_{{\cal N}_{N-1}=0}^{q-\sum_{i=1}^{N-2}{\cal N}_i}\nonumber\\
&&\lambda_1^{{\cal N}_1}\lambda_2^{{\cal N}_2}\ldots
\lambda_{N-1}^{{\cal N}_{N-1}}\lambda_N^{q-\sum_{i=1}^{N-1}
{\cal N}_i}\ .
\label{mq47}
\end{eqnarray}
To end, we will show by induction that for any positive integer $q$
the quantity $\mu_{q,N}$ can be written as
\be
\mu_{q,N}=\hspace{-.2cm}\sum_{{\cal N}_1=0}^q
\sum_{{\cal N}_2=0}^{q-\nu_1}\ldots\hspace{-.4cm}
\sum_{{\cal N}_{N-1}=0}^{q-\sum_{i=1}^{N-2}{\cal N}_i}\hspace{-.2cm}
\lambda_1^{{\cal N}_1}\lambda_2^{{\cal N}_2}\ldots
\lambda_{N-1}^{{\cal N}_{N-1}}\lambda_N^{q-\sum_{i=1}^{N-1}
{\cal N}_i}\ .
\label{assumption}\ee
As a starting point for this line of reasoning we are going to
demonstrate that $\mu_{-1,N}=0$.
For $N=2$ this equality can be checked by direct calculation.
For $N>2$ we have
\begin{eqnarray}
\mu_{-1,N}&=&
\sum_{i=1,i\neq j}^N\frac{\lambda_i^{N-3}}
{\prod_{k=1,k\neq i,k\neq j}(\lambda_i-\lambda_k)}
\frac{\lambda_i}{\lambda_i-\lambda_j}+\nonumber\\
&&\frac{\lambda_j^{N-2}}{\prod_{k=1,k\neq j}(\lambda_j-\lambda_k)}\ .
\end{eqnarray}
Since
$\frac{\lambda_i}{\lambda_i-\lambda_j}=
1+\frac{\lambda_\alpha}{\lambda_i-\lambda_j}$,
we can make use of the following relation
$\sum_{i=1,i\neq j}^N\frac{\lambda_i^{N-3}}
{\prod_{k=1,k\neq i,k\neq j}(\lambda_i-\lambda_k)}=
\mu_{q,N-1}(\lambda_1,\ldots,\lambda_{j-1},\lambda_{j+1},\ldots,\lambda_N)$,
in which the right hand side vanishes by assumption.
Therefore we end up with
\be
\mu_{-1,N}=\lambda_j
\sum_{i=1}^N\frac{\lambda_i^{N-3}}
{\prod_{k=1,k\neq i}^N(\lambda_i-\lambda_k)}\ .
\ee
We still have the freedom to choose the index $j$. As the result
does not depend on the choice of $j$, both sides of the equality
have to be zero.
Now we can come back to the proof of (\ref{assumption}).
For $N=2$ it is just a straight forward calculation to show that
\be
\sum_{\nu=0}^q\lambda_1^\nu\lambda_2^{q-\nu}=
\frac{\lambda_1^{q+1}}{\lambda_1-\lambda_2}+
\frac{\lambda_2^{q+1}}{\lambda_2-\lambda_1}=
\mu_{q,N}\ ,
\ee
which can be checked by multiplying both sides with $\lambda_1-\lambda_2$.
Assuming that (\ref{assumption}) is true for $N-1$ we get
\be
m_{q,N}=\sum_{\nu=0}^q \mu_{q-\nu,N-1}\lambda_N^\nu\ .
\ee
Making use of the assumption (\ref{assumption}) one obtains
\begin{eqnarray}
m_{q,N}&=&
\sum_{i=1}^{N-1}\frac{\lambda_i^{N-2}}
{\prod_{j=1,j\neq i}^{N-1}(\lambda_i-\lambda_j)}
\frac{\lambda_i^{q+1}-\lambda_N^{q+1}}{\lambda_i-\lambda_N}
\nonumber\\
&=&\sum_{i=1}^{N}\frac{\lambda_i^{q+N-1}}
{\prod_{j=1,j\neq i}^{N}(\lambda_i-\lambda_j)}-
\nonumber\\
&&\hspace{-1.3cm}-\lambda_N^{q+1}
\sum_{i=1}^{N}\frac{\lambda_i^{N-2}}
{\prod_{j=1,j\neq i}^{N}(\lambda_i-\lambda_j)}
\frac{\lambda_i^{q+1}-\lambda_N^{q+1}}{\lambda_i-\lambda_N}\ .
\end{eqnarray}
Now one just has to make use of $\mu_{-1,N}=0$ and one immediately
gets that the assumption (\ref{assumption}) holds true for $N$.
This result together with (\ref{mq47}) 
completes the proof of Eq. (\ref{mq1}) for integer positive integers
$q$.
Since integers are dense at $\infty$,
we conclude that there is a unique generalisation to real $q$.
Thus eq (\ref{mq1}) is also valid for real $q$.
Eq (\ref{moments})
for the moments of the Husimi function of a monopartite system 
differs by a proportionality constant only 
and its proof is analogous.
 
\section{Integral representation of the moments}

In order to prove that
\be
\mu_{q,N}=
\frac{\Gamma(q+N)}{\Gamma(q+1)}\int_\Delta
dx(\vec\lambda\vec x)^q\ ,
\label{intexp}
\ee
where the integration is performed over the propability simplex $\Delta$,
we consider the function $h_{q,N}(a,b)$ of two real variables $a$ and
$b$ defined as
\begin{eqnarray}
h_{q,N}&=&\frac{\Gamma(q+N)}{\Gamma(q+1)}
\int_0^{1-a}\hspace{-.5cm}dx_{n-1}\hdots
\int_0^{1-a-\sum_{i>1}x_i}\hspace{-1.2cm}dx_1\nonumber\\
&&(\sum_{i=1}^{n-1}x_i\lambda_i+
\bigl(1-a-\sum_{i=1}^{n-1}x_i)\lambda_n+b\bigr)^q\ .
\label{intab}
\end{eqnarray}
Now we are going to show by induction that $h$ can be expressed as
\be
h_{q,N}(a,b)=\sum_{i=1}^n\frac{\bigl((1-a)\lambda_i+b\bigr)^{q+n-1}}
{\prod_{j\neq i}(\lambda_i-\lambda_j)}\ .
\label{intass}
\ee
For $N=2$ it is straight forward to check that the assumption
\ref{intass} holds.
For $N>2$ we obtain
\begin{eqnarray}
h_{q,N}=\frac{\Gamma(q+N)}{\Gamma(q+1)}\sum_{i=1}^{N-2}
\frac{1}{\prod_{j\neq i}(\lambda_i-\lambda_j)}
\int_0^{1-a}dx_{N-1}\nonumber\\
\bigl((1-a-x_{N-1})\lambda_{N-1}+b+x_{N-1}\lambda_{N-1}\bigr)^{q+N-2}\ ,
\end{eqnarray}
where we were using the assumtion for
$h_{q,N-1}(a+x_{n-1},b+x_{n-1}\lambda_{n-1}$.
Performing the integration one gets
\begin{eqnarray}
h_{q,N}=\frac{\Gamma(q+N)}{\Gamma(q+1)}
\left(\sum_{i=1}^{N-2}
\frac{\bigl((1-a)\lambda_i+b\bigr)^{q+n-1}}
{\prod_{j\neq i}(\lambda_i-\lambda_j)}+\right.\nonumber\\
\left.\sum_{i=1}^{N-2}
\frac{\bigl((1-a)\lambda_{N-1}+b\bigr)^{q+n-1}}{(\lambda_{N-1}-\lambda_i)\prod_{j\neq i}(\lambda_i-\lambda_j)}
\right)
\end{eqnarray}
We are done, if we manage to show that
\begin{eqnarray}
\sum_{i=1}^{N-2}
\frac{1}{(\lambda_{N-1}-\lambda_i)
\prod_{j\neq i}^{N-2}(\lambda_i-\lambda_j)}=\nonumber\\
\frac{1}{\prod_{j=1}^{N-1}(\lambda_{N-1}-\lambda_j)}\ ,
\end{eqnarray}
or eqivalently
\be
p=\sum_{i=1}^{N-2}\prod_{j\neq i}^{N-2}
\frac{\lambda_{N-1}-\lambda_j}{\lambda_i-\lambda_j}=1\ .
\ee
This is a polynomial of $(N-2)$-nd order in $\lambda_{N-1}$.
If we find $N-1$ different values for $\lambda_{N-1}$ such that $p$ equals $1$,
the polynomial $p$ has to be identically equal to unity.
Inserting $\lambda_k$ ($k=1,\hdots,N-1$)
one gets
\be
p=\prod_{j\neq k}\frac{\lambda_k-\lambda_j}{\lambda_k-\lambda_j}=1
\ee
Thus $p\equiv 1$, which completes the proof of eq. (\ref{intass}).
Setting $a=b=1$ in eq. (\ref{intab}) and eq. (\ref{intass}), we
immeadiately get eq. (\ref{intexp}).

\bibliography{referenz2}

\end{document}